\documentclass[preprint,aps,draft,showpacs]{revtex4}
\usepackage{graphicx,amsmath}
\date{\today}
\begin{document}
\title {Excitonic Photoluminescence in Semiconductor Quantum Wells:
Plasma versus Excitons}
\author{S.~Chatterjee}
\author{C.~Ell}
\author{S.~Mosor}
\author{G.~Khitrova}
\author{H.~M.~Gibbs}
\affiliation{Optical Sciences Center, The University of Arizona,
Tucson, Arizona 85721-0094}
\author{W.~Hoyer}
\author{M.~Kira}
\author{S.~W.~Koch}
\affiliation{Department of Physics and Materials Sciences Center,
Philipps-University Marburg, Renthof 5, 35032 Marburg, Germany}
\author{J.~P.~Prineas}
\affiliation{Department of Physics and Astronomy, University Iowa,
Iowa City, Iowa}
\author{H.~Stolz}
\affiliation{Department of Physics, University of Rostock,
Universit\"{a}tsplatz 3, D-18051, Rostock, Germany}
\date{\today}
\renewcommand{\baselinestretch}{1.5}
\newcommand{\be}{\begin{equation}}
\newcommand{\ee}{\end{equation}}
\newcommand{\bea}{\begin{eqnarray}}
\newcommand{\eea}{\end{eqnarray}}
\newcommand{\eps}{\varepsilon}
\newcommand{\ev}[1]{\langle {#1} \rangle}
%

\begin{abstract}
Time-resolved photoluminescence spectra after nonresonant
excitation show a distinct 1s resonance, independent of the
existence of bound excitons. A microscopic analysis identifies
excitonic and electron-hole plasma contributions. For low
temperatures and low densities the excitonic emission is extremely
sensitive to even minute optically active exciton populations making it possible to
extract a phase diagram for incoherent excitonic populations.
%
\end{abstract}
%
\maketitle

For a long time, photoluminescence (PL) at the spectral position
of the 1s exciton resonance has been considered as evidence for
the existence of excitons. The rise of the 1s PL after nonresonant
excitation of a semiconductor was interpreted as buildup of an
excitonic
population~\cite{kusano,blom,eccleston,damen,kumar,gulia}, and the
PL decay was used to describe exciton
recombination~\cite{feldmann,deveaud}. However, recently a
microscopic theory predicted that PL at the 1s resonance can also
originate from correlated plasma emission~\cite{kira98}.
Accordingly, PL at the spectral position of the 1s
resonance would not prove the existence of excitons, and previous
interpretations may be in question. Indeed, in
nonresonantly excited time resolved PL measurements the 1s
resonance is developed on a sub-ps timescale at
100\,K~\cite{hayes}, much faster than any expected exciton
formation time.

Information about exciton formation can be gained by performing
THz experiments \cite{kira01}. However, currently the THz results
are inconclusive: Kaindl et al.~\cite{kaindl} observed the
build-up of the induced absorption corresponding to the excitonic
1s to 2p transition showing excitonic populations with formation
times on a rather slow timescale of 100's of ps to ns. They claim
to observe a nearly completely excitonic system 1\,ns after
excitation, while Chari et al.~\cite{chari} only plasma
contributions. Also, THz absorption is sensitive to both dark and
bright excitons, and cannot answer if and how excitonic
populations influence the PL. Here we address the following
questions. \textit{Is the 1s PL ever dominated by plasma emission?
If so, is it always dominated by plasma emission, i.e. what can we
learn about excitonic populations from the 1s PL and nonlinear
absorption?}

After ps continuum excitation, time resolved PL and corresponding
probe absorption measurements are performed under identical
conditions on a ns timescale. The sample (DBR42) consists of 20
MBE-grown 8\,nm In$_{0.06}$Ga$_{0.94}$As quantum wells with
130\,nm GaAs barriers;
both sides are anti-reflection coated. This indium concentration
places the 1s exciton resonance at 1.471\,eV at 4\,K,
avoiding absorption in the bulk GaAs substrate that leads to
impurity emission at 1.492\,eV from unintentional carbon in the
substrate. The results, checked on several other samples including
a sample grown in a different MBE system,
are insensitive to exciton linewidths or to
interfacial or alloy disorder. Single-quantum-well data
were noisier but exhibited a similar behavior, excluding
significant radiative coupling effects. We excite
nonresonantly 13.2\,meV above the 1s resonance, into the
heavy-hole continuum but below the light-hole resonance. The laser
pulses are generated by a Ti:sapphire oscillator emitting 100\,fs
pulses at 80\,MHz.
Because of the long PL lifetimes, we use a pulse picker to reduce the repetition rate
to 2\,MHz and sweep slowly the
Hamamatsu streak camera, decreasing the PL time resolution to 90 ps.
For spectrally selective excitation, a tunable 3\,ps pulse is
generated in a pulse shaper.
The pump spot is focussed to 60\,$\mu$m diameter, three times
larger than the probe spot. Both absorption and PL are collected
in transmission geometry and spectrally resolved with identical
grating monochromators with a spectral resolution of 0.8\,meV
\cite{filter}. The probe absorption is detected with a liquid
nitrogen cooled Si charge-coupled device. The carrier densities at
different time delays are estimated from a calibration curve
showing the peak height of the nonlinear absorption at 10\,ps
versus the initial carrier density.

Figure~\ref{expspectra} displays measured PL spectra 1\,ns after
nonresonant excitation for three different densities (solid). All
PL spectra exhibit a distinct peak at the 1s exciton resonance
(0\,meV). The second peak 6.5\,meV above is due to emission at the
2s resonance; the energetic separation of the resonances yields an
8\,meV binding energy. The continuum PL exhibits an exponential
fall-off towards higher energies, from which a carrier temperature
is extracted via the Boltzmann factor $\exp(-\Delta E/k_BT)$ by a
least squares fit. In order to get a small standard deviation for
the temperature values, it is necessary to detect several meV of
continuum emission; this requires four to five orders of magnitude
dynamic range.

Carrier temperatures extracted from the spectra shown in
Fig.~\ref{expspectra} are density dependent and vary from 13.5\,K
at the lowest density to 20.7\,K at the highest density. Even at
later times, they never reach the lattice value of
4\,K~\cite{leo,yoon,schnabel}; the lowest temperature measured was
10.5$\pm 0.5$\,K (2.8\,ns after excitation for an initial carrier
density of $n_{\rm eh}=2.9\times 10^8$\,cm$^{-2}$). For a lattice
temperature of 50\,K, our measurements show identical lattice and
carrier temperatures after 0.1\,ns.

To compare many spectra as a function of density and temperature,
it is convenient to define a single parameter $\beta$ to
characterize each spectrum. In thermodynamic equilibrium, one
expects the PL to be proportional to the absorption coefficient
$\alpha$ times a Bose distribution function $g(\hbar\omega-\mu$) =
1/($\exp^{(\hbar\omega-\mu)/k_BT}-1)$, i.e.~$I_{\rm PL}^{\rm
eq}(\hbar\omega) \propto g(\hbar\omega-\mu)) \alpha(\hbar\omega)$
where $\mu$ is the joint chemical potential of the electron-hole
plasma; this is known as the Kubo-Martin-Schwinger (KMS) relation
\cite{kubo}. We then define $\beta$ = $I_{\rm PL}(1s)$/$I_{\rm
PL}^{\rm eq}(1s)$ as in~\cite{schnabel}. $I_{\rm PL}^{\rm
eq}(\hbar\omega)$ is found by multiplying the measured nonlinear
$\alpha$ by a Boltzmann factor to approximate the Bose function,
using the temperature extracted from the measured continuum
emission and normalizing it to agree with the measured continuum
PL; see dashed lines in Fig.~\ref{expspectra}. Thus
$\beta$ quantifies how the 1s emission of a given spectrum differs
from that expected from the measured absorption assuming
validity of the KMS relation.

Figure~\ref{betaexp-theo} displays the density dependence of
$\beta$ for lattice temperatures of 50\,K (top) and 4\,K(bottom).
We find significant deviation from
$\beta=1$ for all densities, temperatures, and times,
implying that KMS is never valid for the configurations
studied here. The 50\,K
result varies only slightly across the investigated density range.
For the lower lattice temperature, we find pronounced deviations
from the thermal equilibrium result particularly for low
densities. For elevated densities, $\beta$ exhibits an increase to
values of around 0.5.

In order to analyze the experimental
observations, we apply our microscopic theory that treats Coulomb
interacting electrons and holes, phonons and a quantized light
field. The details of the theory can be found in previous
publications \cite{kira98,hoyer:03}. This theory, evaluated at the
level of a Hartree-Fock approximation first predicted
luminescence at the exciton energy without exciton populations%
\cite{kira98}. Meanwhile we have extended the analysis to include
also electron-hole correlations and bound excitons. We use an
adiabatic treatment of the photon-assisted polarizations such that
the steady-state luminescence can be obtained from
%
\begin{multline}
I_{\rm PL}(\omega) = \frac{|d_{\rm cv}^2|\omega}{\eps_{\rm bg}}
{\rm Im}\Bigl[
     \sum_{\nu} \frac{\phi^{r}_{\nu}(r=0)}{E_{\nu} - \hbar\omega - i \gamma_{\nu}}\\
        \sum_{k,k'} (\phi_{\nu}^{l}(k))^\star\, \ev{a^{\dagger}_{c,k'} a_{c,k} a_{v,k'} a^{\dagger}_{v,k} }
\Bigr] ,
\label{eq:lumi1}
\end{multline}
%
where $a^{\dagger}_{\lambda,k}$ creates an electron in band
$\lambda = c,v$ in quantum state $k$. The prefactor in Eq.~(\ref{eq:lumi1})
is determined by the square of the dipole matrix element $|d_{\rm
cv}|^2$ and the background dielectric constant $\eps_{\rm bg}$
\cite{param}.
Equation~(\ref{eq:lumi1}) is reminiscent of the famous Elliott
formula for bandgap absorption \cite{elliott}; it contains a sum
over excitonic states, and the resonances of the denominator show
that the PL peaks at the same excitonic energies as the
absorption. In contrast to the absorption, however, the strength
of the PL is not only determined by the exciton
wavefunctions but also by the source term $\sum_{k,k'}
(\phi_{\nu}^{l}(k))^\star\, \ev{a^{\dagger}_{c,k'} a_{c,k}
a_{v,k'} a^{\dagger}_{v,k} }$. This source contains a
plasma contribution $[\phi^{r}_{\nu}(r=0)]^\star\, \sum_k
|\phi^{l}_{\nu}(k)|^2 f^e_k f^h_k$ which is always present as soon
as electrons and holes are excited. Moreover, the source term
can also describe incoherent bound excitonic correlations $N_{\rm
X}$ which may or may not be in the system.

The theory-experiment comparison over the experimentally relevant
density regime requires the proper description of the microscopic
Coulomb scattering which is different for the correlation terms in
comparison to the scattering of the photon-assisted polarizations.
The exciton basis used in the calculations is not the usual
low-density one, instead, left- and right handed basis functions
$\phi^{l/r}$ have to be taken into account. Furthermore, not only
the eigenenergies $E_{\nu}(\omega)$ but also the broadenings
$\gamma_{\nu}(\omega)$ depend on the excitonic index and the
emission frequency. Separating the excitonic and plasma
populations we evaluate our theory and determine the $\beta$
factor. Here, the respective excitonic and plasma populations are
treated as input to the theory, and a fit to the measured PL
spectra is obtained by varying the optically active ($q = 0$)
1s-exciton population. Thus, this formulation of the theory allows
us to formally distinguish between electron-hole plasma and bound
excitons as different possible sources to excitonic PL.

Computed spectra are shown in Fig.~\ref{theospectra}. There the
pure plasma PL (dotted line) is compared to the KMS result (dashed
line) obtained from the computed absorption. The corresponding
$\beta$ factors are plotted as dotted lines in
Fig.~\ref{betaexp-theo}. At 50\,K, the bare plasma emission shows
good agreement with the experimental $\beta$'s over almost the
whole density range. Even for 4\,K, the highest density results
agree. Note that the exciton absorption is still very pronounced
with the peak reduced by only 25\,$\%$. This answers our first
question: \textit{the 1s PL can be dominated by plasma emission,
and the spectra explained by a pure plasma theory; this is the
case at high temperatures and even at low temperatures for higher
densities.}

Figure~\ref{betaexp-theo} also shows that at 4\,K and low to
intermediate densities, the pure plasma calculation underestimates
$\beta$. We attribute the stronger measured 1s emission to the
presence of incoherent bound 1s-exciton populations $N_{\rm X}$
and test this hypothesis by adding optically active $q=0$ excitons
in the theory. Fig.~\ref{theospectra} shows that this addition
strongly enhances the 1s PL. Since the plasma contribution changes
quadratically with the carrier density, even minute exciton
fractions result in a strong enhancement of the 1s resonance. The
addition of excitons allows us to obtain an excellent
theory-experiment agreement for $\beta$ for all densities; see
solid lines in Fig.~\ref{betaexp-theo}. The inset shows the bright
excitons necessary for that agreement.

When the 1s-exciton distribution is expressed as $N_X(q) =
N_X(q=0)\,F(q)$ with $F(q=0)=1$, only the $q=0$ value can be
deduced from the theory-experiment comparison. Thus it is
impossible to determine the total number of excitons from PL
measurements. If we assume
a thermal distribution at 4\,K, the largest value in the inset of
Fig.~2 corresponds to a total exciton fraction of 4\,\%. However,
if significant hole burning~\cite{piermarocchi,kira01} in the 1s
distribution around $q=0$ is present, this number could be much
larger. This answers our second question: \textit{at low
temperatures and for low and intermediate densities, the 1s PL is
dominated by the recombination of excitons, even though the
density of optically active excitons may be only a small fraction
of the plasma density. Consequently, depending upon temperature
and density, excitonic PL can be almost entirely from the plasma
or almost entirely from excitons.}

Even though our PL studies cannot determine the total exciton
number or the detailed exciton distribution function, our results
clearly show an increased importance of excitons at lower lattice
temperatures and intermediate densities. This observation is in
qualitative agreement with exciton formation studies in quantum
wires~\cite{hoyer:03}. Due to kinetic arguments, exciton formation
after nonresonant excitation should be most efficient at
intermediate densities. At low densities the probability for
electron-hole collisions is too small, whereas at higher densities
exciton binding is hampered by phase space effects and screening
of the attractive Coulomb interaction.

We can further investigate the low-density behavior by studying
the temporal evolution of $\beta$ shown in Fig.~\ref{betatime}.
Because the experiment is pulsed, the carrier density decays with
time. While $\beta$ does not vary significantly in the strong
excitation regime, implying that a quasi-steady-state has been
reached, it decays monotonically at lower densities. The inset,
like Fig.~\ref{betaexp-theo} for low densities, contains $\beta$
values at several time delays all $\geq$ 1\,ns to exclude cooling
effects~\cite{leo}. The functional behavior
follows an interpolated curve through the 1\,ns values (squares).
Here the decay of $\beta$ and the corresponding exciton fraction
are thus fully parameterized by the density decay. This shows that
$\beta$ depends not upon past history, but only upon the momentary
carrier density and temperature, as needed for comparison with
a quasi-steady-state theory.

In conclusion, by carefully mapping out the parameter space of
carrier density and lattice temperature, we have identified
conditions under which the PL emission at the 1s resonance after
nonresonant excitation into the continuum is dominated
by the plasma or by an incoherent excitonic
population. We observe that the appearance of the 1s-exciton
resonance in PL is ubiquitous, independent of the existence of
bound excitons and that the KMS relation is never fulfilled.
In particular, the 1s emission is always weaker than
that predicted by KMS. Under suitable conditions excitons
may form after
nonresonant excitation, and we identify a regime of low
temperature and intermediate density as most favorable.
\begin{acknowledgments}
The work is supported in Tucson by NSF (AMOP), AFOSR (DURINT), and
COEDIP and in Marburg by the Deutsche Forschungsgemeinschaft
through the Quantum Optics in Semiconductors Research Group, by
the Humboldt Foundation and the Max-Planck Society through the
Max-Planck Research prize, and by the Optodynamics Center of the
Philipps-Universit{\"a}t Marburg.
\end{acknowledgments}
%

%
\pagebreak
\begin{figure}
\caption{Experimental PL spectra (solid) for a lattice temperature
of 4\,K are compared to the KMS result $I_{\rm PL}^{\rm
eq}(\hbar\omega)$ (dashed), i.e.~the measured nonlinear $\alpha$L
multiplied by the Boltzmann factor. PL is integrated from 0.95 to
1.05\,ns after nonresonant excitation, and the densities are for t
= 1 ns. The corresponding curve pairs are vertically offset by two
decades. Inset: Nonlinear absorption spectra for the highest
(dashed) and lowest densities (solid).}
\label{expspectra}
\end{figure}
%
%
\begin{figure}
\caption{$\beta$ versus carrier density. Full squares refer to
experimental values taken at 1\,ns after nonresonant excitation at
a lattice temperature of 4\,K. The densities refer to densities at
1\,ns. The theoretical values (dotted line) are calculated using the PL
formula (\ref{eq:lumi1}) for a pure e-h plasma with a carrier
temperature of 16$\pm2$\,K. The solid line shows a theoretical fit
including a $q=0$ exciton contribution. Top: same for 50\,K
lattice temperature and extracted e-h plasma temperature of
50$\pm3$\,K; the theoretical curves are computed for a carrier
temperature of 48\,K. Inset: Phase diagram of $q=0$ excitonic
contributions as a function of the carrier density.}
\label{betaexp-theo}
\end{figure}
%
%
\begin{figure}
\caption{Theoretical PL spectra for various densities for a
carrier temperature of 16 K: pure plasma (dotted) and including
excitons (solid).  The exciton contributions used are chosen to
give agreement with the experimental $\beta$ as shown in
Fig.~\ref{betaexp-theo}. The KMS result (dashed) is obtained by
multiplying the computed nonlinear $\alpha$L by the Boltzmann
factor.} \label{theospectra}
\end{figure}
%
%
\begin{figure}
\caption{Experimental $\beta$ values versus time for 4 K lattice
temperature. For nonresonant excitation $\beta<1$, i.e.~the 1s
emission is less than for thermal equilibrium. The carrier
densities (cm$^{-2}$) at 1\,ns are 2.0$\times$10$^{10}$ (dots),
1.1$\times$10$^9$ (triangles), 4.4$\times$10$^8$ (diamonds), and
3.1$\times$10$^7$ (stars). The inset shows the low density part of
beta versus carrier density, and an additional set (open circles)
with a carrier density of 4.7$\times$10$^7$ at 1\,ns. The data
points shown are taken for times $\geq$ 1ns.}
\label{betatime}
\end{figure}
%
%

\begin{thebibliography}{99}
\bibitem{filter}To better utilize the dynamic range of the streak camera a
film neutral density filter was placed at the exit plane of the
corresponding monochromator reducing the fluence by 40 for all
energies $<$ 1.474\,eV. The shown data are multiplied by this
factor to present them undistorted.
\bibitem{param} Our calculations are performed for finite
quantum well widths and infinite barrier heights. We adjust the
effective well width such that our calculations reproduce the
experimentally measured exciton binding energy. $m_h/m_e = 3$ as determined
using $k \cdot p$ theory, the other material parameters cancel out
of the $\beta$ calculations.
\end{thebibliography}
\end{document}